# Mass Balance Approximation of Unfolding Improves Potential-Like Methods for Protein Stability Predictions

*Ivan Rossi[1], Guido Barducci[1], Tiziana Sanavia[1], Paola Turina[2],*
*Emidio Capriotti[2,\*], Piero Fariselli[1,\*]*

[1] Department of Medical Sciences, University of Torino, Via Santena 19, 10126 Torino, Italy
[2] Department of Pharmacy and Biotechnology (FaBiT), University of Bologna, Bologna, Italy
\* Corresponding authors: emidio.capriotti@unibo.it, piero.fariselli@unito.it

## Abstract

The prediction of protein stability changes following single-point mutations plays a pivotal role in computational biology, particularly in areas like drug discovery, enzyme reengineering, and genetic disease analysis. Although deep-learning strategies have pushed the field forward, their use in standard workflows remains limited due to resource demands. Conversely, potential-like methods are fast, intuitive, and efficient. Yet, these typically estimate Gibbs free energy shifts without considering the free-energy variations in the unfolded protein state, an omission that may breach mass balance and diminish accuracy. This study shows that incorporating a mass-balance correction (MBC) to account for the unfolded state significantly enhances these methods. While many machine learning models partially model this balance, our analysis suggests that a refined representation of the unfolded state may improve the predictive performance.

Availability: The Python codes and the data used in this study can be downloaded from Github at https://github.com/compbiomed-unito/ddMBC

# Introduction

Predicting protein stability changes upon single-point mutations is a longstanding challenge in computational biology[1–3], with significant implications in drug design, enzyme engineering, and understanding disease mechanisms[4]. Protein stability is typically quantified by measuring the Gibbs free energy change (ΔG) between the folded and unfolded states as

$$\Delta G = G_F - G_U \qquad [1]$$

However, mutations can dramatically alter this delicate balance. Destabilizing mutations are often linked to diseases[5,] such as cancer[6], while stabilizing mutations can enhance protein function and resilience, especially in industrial and therapeutic settings[7,8].

From the experimental point of view, the measure of interest is the difference of the unfolding free energy between the mutated and wild-type proteins (ΔΔG), calculated as

$$\Delta\Delta G = (G_F(m) - G_F(w)) - (G_U(m) - G_U(w)) \qquad [2]$$

where m and w stands for mutant and wild-type (Fig. 1)

$$P_F(w) + P_U(m) \rightleftharpoons P_F(m) + P_U(w) \qquad [3]$$

Where *P* represents the concentration of the protein either in the wild-type (w) or mutant (m) forms both in the folded (*F*) or unfolded (*U*) states. It can be noticed that this kind of "reaction" corresponds to that used in Free-Energy Perturbation (FEP) calculations[9,10], a widely-used method to calculate ΔG differences in molecular modeling and drug design.

The folding free energy difference between two protein variants depends on both the folded and unfolded states of each sequence. Studies using molecular dynamics, based on *Alchemical Free Energy Perturbation*[10,11], have demonstrated that accurately modeling the unfolded state is crucial for achieving high predictive performance[11], though such approaches require computationally expensive methods. Similar statistical-mechanics approaches describing the contribution of the unfolded-state have been presented by Bastolla and coworkers[12–14].

In recent years, deep learning-based approaches have significantly advanced the field of protein stability prediction. Despite their success, these models require substantial computational resources and are sometimes inaccessible for routine or high-throughput applications[2].

In contrast, potential-like methods, such as those utilizing empirical energy functions like FoldX[15] structure-based protein-language models such as ProteinMPNN[16] and ESM-IF1[17], and methods that directly address the calculation of ΔΔG upon mutation using deep neural networks, such as Pythia[18], offer faster and more accessible alternatives. These methods estimate stability changes by calculating either atomistic interactions or the likelihood of an amino acid in a given structural context of the protein. Pythia, for example, employs a self-supervised learning framework to perform zero-shot ΔΔG predictions across a large protein sequence space, offering ultrafast computational performance.

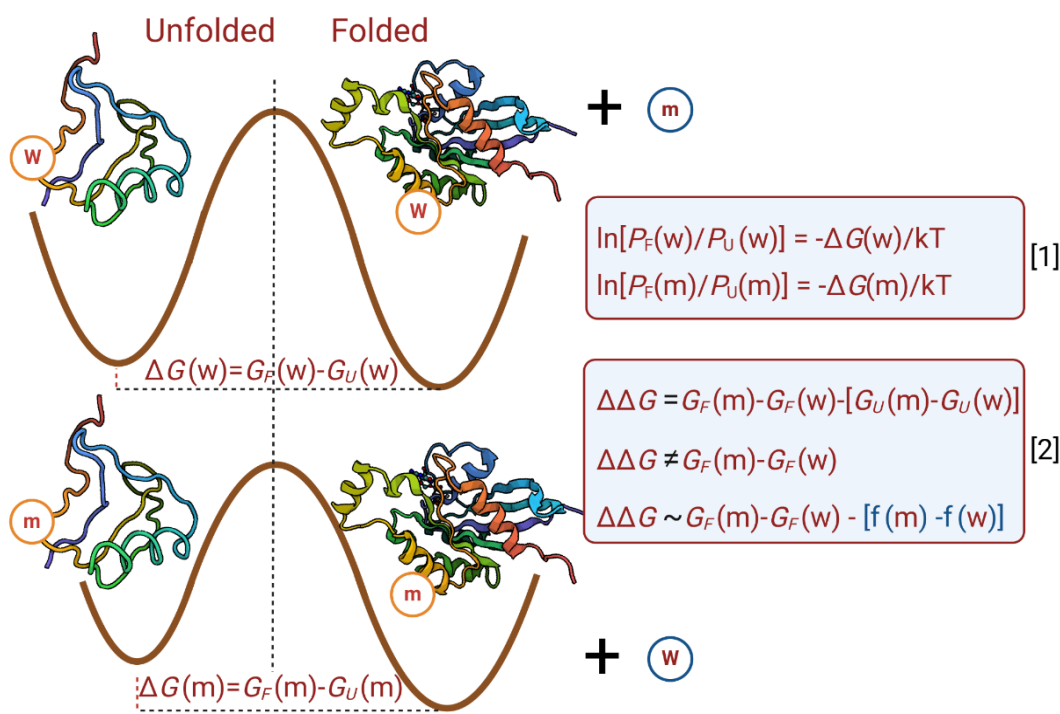

Fig.1 Thermodynamics of the Variation of the folding free energy upon single point mutation, considering mutated (m) and wild-type (w) states. In box [1] the relation between probability and free energy of folding is reported. In box [2], the correct measure of the difference of the unfolding free energy between the mutated and wild-type proteins, considering the difference between the folded and unfolded state is reported (first equation); however, some potential-like methods approximate it using the difference of the folding state free energy, neglecting the effect of the unfolded states (box [2], second equation). A first approximation can be obtained by adding a mass-balance correction (also a kind of solvation term) to the folding free energy difference (box [2], third equation).

However, one fundamental limitation of the potential-like methods is their simplified approach to Gibbs free energy calculations, where only the folded states $\{G_F(x)\}$ (i.e., the protein structure) are considered. This simplification leads to the following approximation for the mutant (m) and wild-type (w):

$$\Delta\Delta G = (G_F(m) - G_F(w)) \quad [4]$$

Under this approximation, the second term of Eq. 2, describing the ΔG between the unfolded states of the two protein sequences, is typically neglected due to the difficulty of properly defining and measuring it. However, this approximation might not always hold, since, for example, different inter-residue interactions and degrees of freedom between wild-type and mutant might persist in the unfolded state. An additional contribution might be the difference in free-energy of solvation for the amino acids involved in the mutation[26]. It should also be observed that the ΔΔG expression is a difference between two terms, and neglecting one could lead to significant deviations from the correct solution. Furthermore, neglecting this second term also implicitly means violating the mass-conservation for the process, as Gibbs free energy is defined for closed systems where mass is conserved.

Considering the extreme flexibility of the neural-networks in implicitly modeling all terms of Eq. 1, the approximation of ΔG between the unfolded states of the two protein sequences equal to zero should not affect, in principle, models that explicitly incorporate the protein-sequence composition change among their input features (e.g. I-mutant[19], ACDC-NN[20], Stability Oracle[21]). However, as previously mentioned, this approximation might become relevant for models that do not compensate for it, such as most "potential-like" methods.

To address this gap, we propose a novel correction that incorporates "mass balance" back into potential-like scoring methods, improving the accuracy of protein stability predictions without compromising their usually high computational efficiency. By retrofitting these potential-like models with this extra term, which we call *mass-balance correction* (MBC), our approach adjusts for a key flaw in the evaluation of ΔΔG, significantly enhancing the prediction accuracy without any reparameterization of the original model.
Furthermore, the obtained performance for some of these modified methods are comparable, or even better, to those of state-of-the-art models such as Stability Oracle, providing a valuable tool for researchers needing rapid stability assessments.

## Results

*Incorporating Mass-Balance Information as a First Approximation of the Unfolded State*

We first evaluated the performance of three different potential-like methods, representing three different approaches to ΔΔG calculation, with and without the MBC correction. Then we compared them to the results of the DDGun3D[22,23]

"untrained" benchmark model. DDGun3D explicitly incorporates a form of MBC by considering the hydrophobicity difference between mutated and wild-type residues, establishing it as a suitable reference benchmark. We also derived the data-driven MBC term, referred to as MBC(dd) hereafter, by fitting it to the training set using ridge regression implemented in Scikit-learn[24] with default parameters.

The MBC(dd) term was then compared with the Kyte-Doolittle[25] and Rose[26] scales to score the difference between hydrophobicity and solvation, respectively, as first approximations of the unfolded state. Additionally, we included a comparison with the Stability Oracle model, a recent state-of-the-art deep learning-based method. We used the S461 dataset[27] as the test set to perform comparisons.

The three potential-like methods considered are:
1. **ESM-IF1**, a large protein-language model (PLM) trained to predict a protein sequence likelihood from its backbone atom coordinates;
2. **FoldX**, a widely-used all-atom knowledge-based potential for fast and quantitative estimation of the importance of the interactions contributing to the stability of proteins;
3. **Pythia**, a self-supervised graph neural network tailored for zero-shot ΔΔG predictions, large-scale residue scanning and missing-residue probability prediction.

On the S461 test set, all methods showed visible performance boosts, with increased Pearson correlation coefficients (PCC) compared to the original methods and with Pythia/MBC(dd) being the top-performer.

Although we used PDB structures to train our model, we observed that the performance of both the baseline ESM-IF1 and Pythia models noticeably depends on the type of structure used. Namely, the performance of both of these methods is higher if AlphaFold[28] models are used instead of experimental X-ray structures from PDB. This is probably due to the way these methods have been parameterized: for both ESM-IF1 and Pythia training sets, the percentage of AlphaFold structure exceeds 90%, thus any bias that may be introduced by using models instead of experimental structures is captured by the methods. Nonetheless, the MBC(dd) validity is not affected by the choice of the model origin (Table S2): using the MBC(dd) correction derived from the PDB structures on the same test sets, but giving in input the AlphaFold structures, instead of those from PDB, results in models that are even better-performing. Both Stability Oracle and Pythia/MBC(dd)-AF achieve a PCC higher than the one obtained by the benchmark DDGun3D method (PCC: 0.62), whose performance on the S461 data set is very strong (Figure 2). We also computed the MBC(dd) correction for Stability Oracle and DDGun3D benchmarks, and, as expected, the result is worse for both methods (Figuure 2). This supports our expectation that these methods, which already account for descriptors of the unfolded state in their input, such as the stoichiometry of the mutation process, are effectively capturing the correct underlying physics without requiring any posterior corrections.

*Comparison between residue specific-coefficients and experimental solvation scales*

We performed a Pearson correlation analysis among the residue-specific parameters fitted using the VBS3322 dataset (see Methods section) to assess their consistency across different methods. Additionally, we included solvation and hydrophobicity scale values in the correlation comparison to evaluate their relationship with the fitted parameters. As shown in Fig. 3, the amino acid-specific parameters ($a_1$ to $a_{20}$) exhibit strong correlations across the potential-like methods. Furthermore, these fitted parameters show a notable correlation with Kyte and Doolittle hydrophobicity scale and an even stronger correlation with the experimentally-derived Rose scale, which was specifically designed to predict the average change in solvent accessible surface area of amino acids upon folding.

In agreement with these observations, we then computed a new MBC based on the Rose scale, referred to as MBC(Rose). This correction was derived using a two-parameter linear combination between the original-method delta and the Rose-scale delta (see Eq. 8), with results summarized in Fig. 2. The performance of MBC(Rose) is consistent with, or in some cases superior to, that obtained by the MBC(dd) approach.

As a further validation, we computed the Pythia/MBC(dd) and Pythia/MBC(Rose) scores using the parameters derived from our VBS3322 training set and tested them on the independent mega-scale dataset[29], which was not used in the parameter derivation. The results show an improvement (PCC: +0.07) over the original Pythia score, achieving a PCC close to 0.70 and an RMSE of 1.43 kcal/mol.

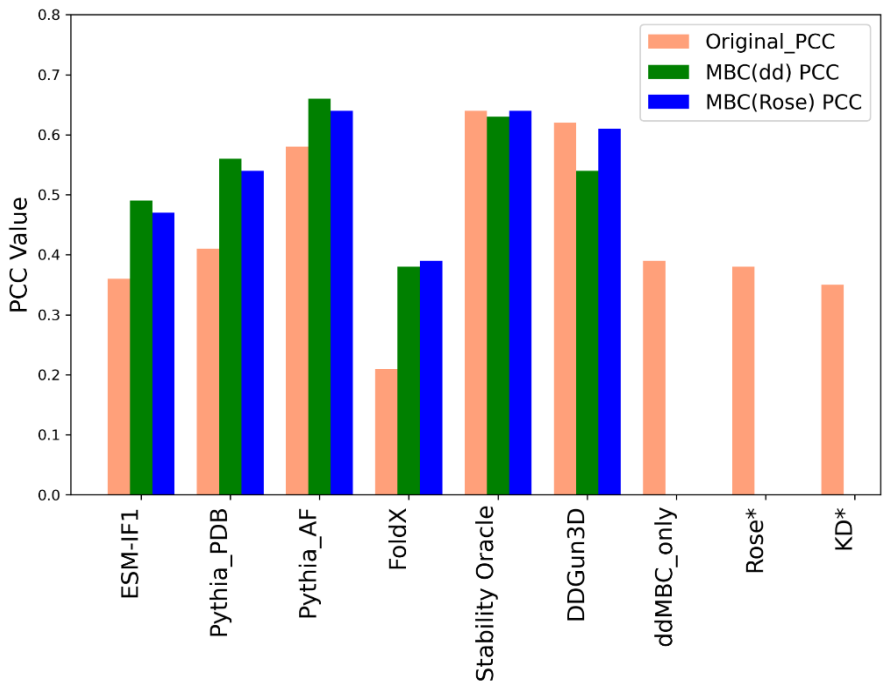

Fig.2 Comparison of Pearson correlation Coefficient obtained on S461 dataset between the original method (pink bar) and its adjusted version with Mass-Balance Correction, using bothMBC(dd) (green bar) and Rose scale (blue bar). ddMMBC_only represents the prediction made using only the fitted mutation coefficients without incorporating a method. *KD[25] and *Rose[26] and are the scale difference values without any fitting.

*Generalization of the Mass-Balance Correction Across Different Methods*

Reeves and Kalyaanamoorthy[30] recently highlighted that structure-based and sequence-based PLMs can be linearly combined to improve the performance, indicating that these two methodological classes provide complementary information. They further noted that "...*PSLMs can be reliably augmented with physicochemical properties to exceed the median performance of the benchmark stability predictor..*". This aligns with our model, since

$$\Delta\Delta G = (G_F(m) - G_F(w)) - (G_U(m) - G_U(w)) = \Delta\Delta H_F - T\Delta\Delta S_F \quad [5]$$

Thus, it is reasonable to think that both sequence and structure-based terms correspond to the ΔΔG term for the unfolded and folded states, respectively.
Additionally, the molecular volume and the solvent-accessible surface area (SASA) play a crucial role in estimating the solvation energy changes (a large part of $\Delta\Delta S_F$) when a molecule interacts with a solvent. This concept has been widely applied in different implicit solvation models, such as the GBSA family of models[31,32].
From this perspective, the MBC can be seen as a proxy of this information. Our model provides a simple, yet effective, way to estimate the Gibbs free energy difference between wild-type and mutated proteins in their unfolded states.

Alternatively, it can be interpreted as describing the differences in the entropy of folding (which is largely dictated by solvation effects), while the potential-like methods primarily approximate the enthalpic contribution to the folding.

We thus tested whether our approach is able to generalize across different methods, considering the predictions of 48 methods on S461 dataset taken from Reeves and Kalyaanamoorthy[30] and supplemented by the Pythia data. To fit the two scale values related to the method and to the Rose scale (see Methods equation 8) we used the prediction reported by the same authors on the Ssym dataset[33].

Fig. 4 reports the obtained results, grouping the methods into MBC-aware (i.e. trained with some mass-balance correction) and non-MBC-aware approaches (such as PLMs, which does not account for the mass balance). As expected, the MBC approach notably improved the performance of non-MBC-aware methods.

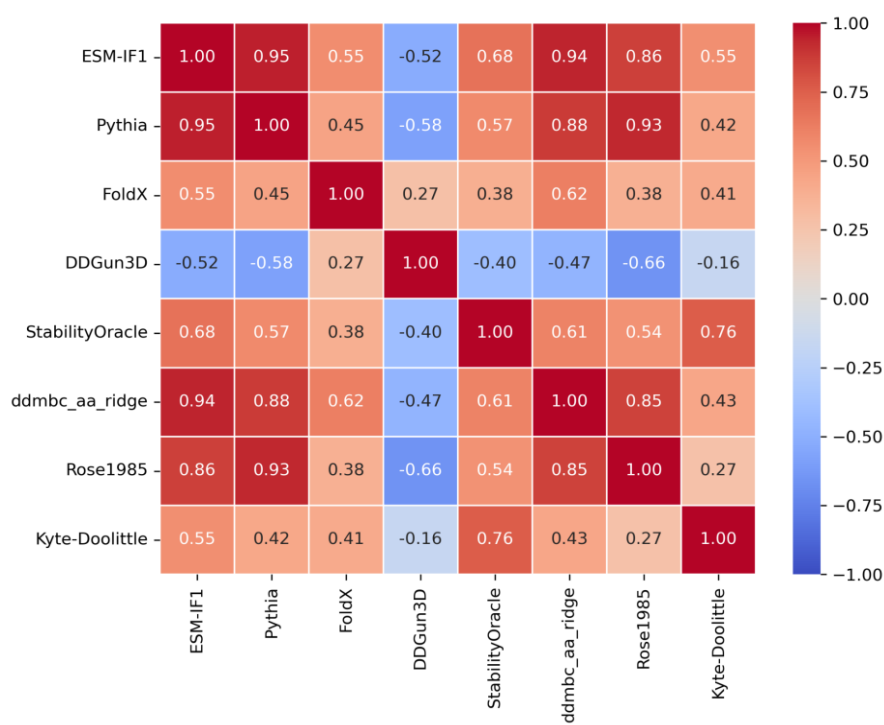

Fig.3 Correlation among the residue coefficients of the different methods and two hydrophobicity scales (Kyte-Doolittle[25] and Rose[26]). DDGun3D contains explicitly the difference of the Kyte-Doolittle values. ddmbc_aa_ridge is highly correlated with the Rose scale

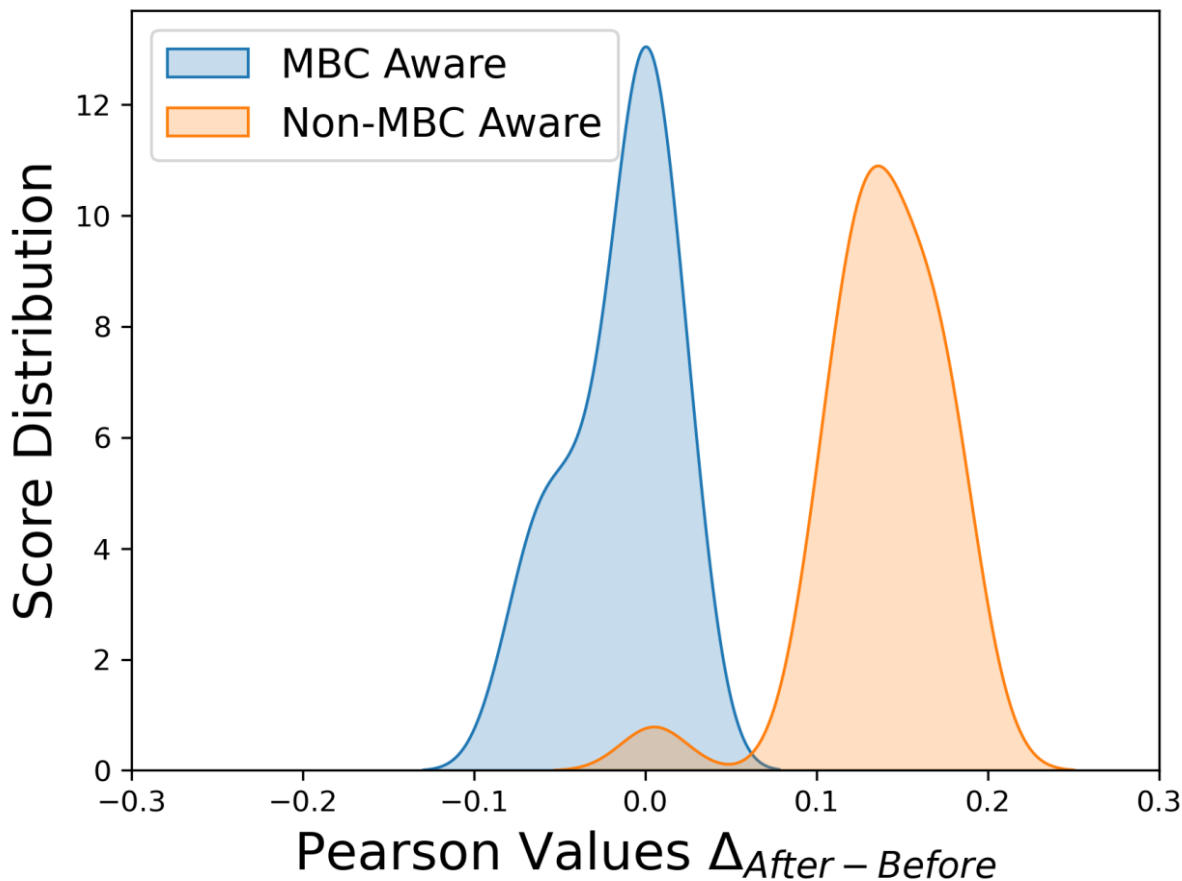

Fig.4 Comparison with methods that directly include a mass-balance correction (MBC Aware) with those that compute only a difference between the folding states (Non MBC Aware). The plot reports the distribution of the difference between the Pearson's correlation after and before the mass-balance term is added. The data are from Reeves and Kalyaanamoorthy[30]

**Conclusions**

The mass-balance correction (MBC), whether data-driven or based on an experimentally derived scale, demonstrates broad applicability, enhancing the performance of various potential-like methods developed through different approaches. These include knowledge-based potentials, sequence- and structure-based protein language models (PLMs), and a self-supervised deep graph-neural network. Notably, MBC achieves these improvements without requiring any re-parameterization of the base methods and with negligible additional computational cost.

In several cases, the enhancement of the performance due to MBC is substantial. Specifically, in the case of Pythia, the results are particularly notable, bringing Pythia-MBC close to state-of-the-art performance while also addressing the method's poor antisymmetry (from -0.53 to -0.68 of antisymmetry in Ssym). More generally, MBC preserves the antisymmetry of the improved methods whenever the original methods exhibit this property.

This finding strongly supports our hypothesis that a better description of the unfolded state of the proteins might be a necessary step to improve the current state-of-the-art protein stability-change predictions. The MBC correction is just a simple, yet effective, zero-order correction. Thus, it is clearly possible to envision more sophisticated and, eventually, better-performing methods. Nonetheless, we believe that the simplicity of our approach has its own merits per se, since it allows the retrofitting of several existing approaches, achieving good performance and avoiding extra computational costs.

## Materials and Methods

### Datasets composition

The main training set used in this work, namely VBS3322, consists of 3,322 mutations obtained by combining the VariBench[35] and the S2648[34] data sets. In the cases where the same mutation is reported in both data sets, the VariBench value is considered. We also augmented the dataset by including the antisymmetric complement of each mutation, as suggested in a previous work[36].

The test set used for the benchmarking is the S461. For all the structures that showed missing backbone atoms, we preprocessed the structure using the PDBFixer utility[37].

The FoldX results used for both training and evaluation have already been published[38], while Stability Oracle results for the S461 dataset have been computed from the data provided by its authors on Github (https://github.com/danny305/StabilityOracle)

### Mass balance Correction

The simplest approach to calculate the ΔΔG for the sequence-mutation process is to assume that the second term of equation [1] depends only on the amino acids involved in the mutation.

This simplification leads to the following reaction, considering the wild-type ($w$) and mutated ($m$) residues:

$$Protein(w,i) + Residue(m) \rightleftarrows Protein(m,i) + Residue(w) \quad\quad [6]$$

where $Protein(x,i)$ represents a protein with residue $x$ in position $i$, while $Residue(x)$ refers to a single amino acid. Conceptually, this corresponds to estimating the difference in the (effective) Gibbs free energy of solvation for two amino acids in solution and within the field of the protein. From another perspective, this approach approximates the free energy of the unfolded state as the sum of independent contributions from each amino acid. Physically, these contributions may arise from the conformational entropy of both the side chain and main chain, as well as their interactions with the solvent. Under this approximation, all terms disappear except for the contributions of the wild-type and mutated amino acids, significantly simplifying the calculation.

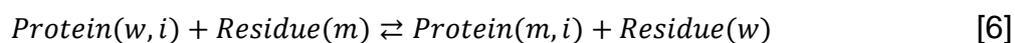

### Input encoding

We encode the mutation in the sequence as a twenty-elements array, one element for each of the natural amino acids, and we encode their occurrence (*O*) as -1 for the wild-type amino acid and +1 for the substitution.

The modified expression to calculate ΔΔG is then expressed as a linear combination of the original-method score (S) for the wild-type and the mutated protein:

$$\Delta\Delta G = a_0(S(m) - S(w)) + \sum_{i=1}^{20} a_i O_i \qquad [7]$$

the first term represents the original method's (scaled) output and the second term represents the pseudo-ΔΔG of solvation for the amino acids involved in the mutation (the data-driven MBC).
The first term $S(x)$ thus corresponds to the ΔΔG predicted by the original method, while the second term depends on amino acid-related parameters.
It should also be observed that equation [7], being antisymmetric by definition, preserves the antisymmetry in the prediction of the original methods, if present.
The 21 coefficients for the linear model above can be easily derived via a simple linear regression with respect to the training set.

Similarly, the MBC(Rose) correction is computed as a two-parameter linear combination of the original-method score (*S*) and Rose-scale delta

$$\Delta\Delta G = a_0(S(m) - S(w)) + a_1(R(m) - R(w)) \qquad [8]$$

where $R(m) \wedge R(w)$ are the values of the Rose scale for the mutated- and wild-type amino acid respectively.

### Measures of performance

To evaluate the performance of the methods in the regression task, we compared the predicted (p) and experimental (e) values of the variation of unfolding free energy change upon mutation (ΔΔG). The standard scoring values calculated in our assessment are the Pearson correlation coefficients (*PCC*) and the root mean square error (RMSE), defined as follows:

$$PCC = \frac{\sum_{i=1}^{N}\left(\Delta\Delta G_e - \overline{\Delta\Delta G_e}\right)\left(\Delta\Delta G_p - \overline{\Delta\Delta G_p}\right)}{\sqrt{\sum_{i=1}^{N}\left(\Delta\Delta G_e - \overline{\Delta\Delta G_e}\right)^2}\sqrt{\sum_{i=1}^{N}\left(\Delta\Delta G_p - \overline{\Delta\Delta G_p}\right)^2}} \qquad [9]$$

$$RMSE = \sqrt{\frac{\sum_{i=1}^{N}(\Delta\Delta G_p - \Delta\Delta G_e)^2}{N}} \qquad [10]$$

where $\overline{\Delta\Delta G_p}$ and $\overline{\Delta\Delta G_e}$ are the average predicted and experimental ΔΔG values, respectively.

## Acknowledgments

The authors thank the Italian Ministry for Education, University and Research under the programme "Ricerca Locale ex-60%" and PNRR M4C2 HPC—1.4 "CENTRI NAZIONALI"- Spoke 8 for fellowship support. In addition, the authors thank the European Union's Horizon 2020 projects Brainteaser (Grant Agreement ID: 101017598) and GenoMed4All (Grant Agreement ID:101017549:). IR would like to dedicate this work to his past supervisor Prof. Donald G. Truhlar.

.

.